\documentclass{svproc}
\usepackage{url}

\usepackage[square,numbers]{natbib}
\begin{document}
\mainmatter              
\title{DSMC: A Statistical Mechanics Perspective}
\titlerunning{DSMC \& Statistical Mechanics}  
%
\author{Alejandro L. Garcia\inst{1}
}
\authorrunning{Alejandro L. Garcia} 
%
%
\institute{San Jose State University, San Jose CA 95192 USA,\\
\email{Alejandro.Garcia@sjsu.edu},\\ WWW home page:
\texttt{http://www.algarcia.org}
}

\maketitle              

\begin{abstract}
This paper presents a perspective in which Direct Simulation Monte Carlo (DSMC) is viewed not in its traditional role as an algorithm for solving the Boltzmann equation but as a numerical method for statistical mechanics. 
First, analytical techniques such as the collision virial and Green-Kubo relations, commonly used in molecular dynamics, are used to study the numerical properties of the DSMC algorithm. 
The stochastic aspect of DSMC, which is often viewed as unwanted numerical noise, is shown to be a useful feature for problems in statistical physics, such as Brownian motion and thermodynamic fluctuations.
Finally, it is argued that fundamental results from statistical mechanics can provide guardrails when applying machine learning to DSMC.
\end{abstract}

\section{MD and DSMC -- Brothers separated at birth}

Three scientific eras dawned in the middle of the last century: the atomic age, the space age, and the age of the computer.
Numerical algorithms were developed for the study of fluids under extreme conditions, such as nuclear explosions and orbital reentry, that were beyond the realm of conventional laboratory experiments.
An early example is the Metropolis Monte Carlo method, which in 1953 showed that a particle representation could be used to map phase space and using statistical mechanics extract a fluid's thermodynamic properties, such as the equation of state.\cite{metropolis1953equation}
It was closely followed by molecular dynamics (MD), introduced in 1959 by Alder and Wainwright, which uses a similar particle representation but includes dynamics, allowing it to measure transport properties such as viscosity.\cite{alder1959studies}

In 1963, Graeme Bird submitted to \textit{Physics of Fluids} a paper entitled ``Approach to Translational Equilibrium in a Rigid Sphere Gas.'' 
In this article he introduced a stochastic collision algorithm that would become the cornerstone of Direct Simulation Monte Carlo (DSMC). 
The paper was rejected (G.A. Bird, personal communication).
The referee had failed to recognize that DSMC differed significantly from MD; after revising it to explain these differences, Bird's paper was accepted.\cite{bird1963approach} 

DSMC went on to become the dominant numerical method in aerospace engineering for rarefied gas dynamics while MD established popularity in the field of statistical mechanics for the simulation of microscopic systems.   
For many years, these communities remained separate, yet recently they have found common interests in topics such as nanofluidics and granular flows.
This paper presents several short stories in which DSMC is used by researchers, such as myself, whose background is in statistical mechanics and molecular dynamics.
As we shall see, by approaching DSMC from an unconventional perspective, a number of new avenues are opened.
This narrative is intentionally personal, since Graeme Bird, to whom this paper is dedicated, was a colleague and friend for nearly 30 years. 
In the interest of space, a summary of the DSMC algorithm is omitted; see~\cite{alexander1997direct} for an introduction and \cite{bird1994molecular,bird2013dsmc,boyd2017nonequilibrium} for full treatments.

\section{But why ideal gas law?}

In 1989 I joined the faculty at San Jose State University, located in Silicon Valley. 
About that time I met Berni Alder and started working with him at the nearby Livermore Lab.
Alder took an immediate interest in DSMC and we eventually published 16 papers together, most of them on ways to extend the DSMC algorithm.
Yet when I first explained DSMC to Berni, he was puzzled and asked ``But why does it give the ideal gas law?''
Graeme Bird did not address this question in his books on DSMC~\cite{bird1976molecular,bird1994molecular,bird2013dsmc} since he always viewed the algorithm as numerically solving the Boltzmann equation, a perspective that was reinforced by a proof by Wagner~\cite{wagner1992convergence}.

In their seminal MD simulations Alder and Wainwright~\cite{alder1960studies} measured the equation of state for hard spheres (mass $m$, diameter $d$) as
\begin{equation}
    p = n k T + {\textstyle \frac13} m \Gamma \Theta
    \label{eqn:IGL}
\end{equation}
where $p$, $n$, $T$ are pressure, number density, and temperature; $k$ is the Boltzmann constant.
The collision rate is $\Gamma$ and
\begin{equation}
    \Theta = \langle \Delta \vec{v}_\alpha \cdot \vec{r}_{\alpha\beta} \rangle \geq 0
    \label{eqn:virial}
\end{equation}
is the virial.
For a collision between particles $\alpha$ and $\beta$, $\Delta \vec{v}_\alpha$ is the change in velocity for particle $\alpha$ and $\vec{r}_{\alpha\beta} = \vec{r}_{\alpha} - \vec{r}_{\beta}$ is their separation.
The virial is an average over collisions and depends on the particle diameter since $|\vec{r}_{\alpha\beta}|=d$.
In (\ref{eqn:IGL}) the pressure is the sum of translational and collisional contributions to momentum transfer.

In DSMC collisions occur within cells and the probability of a collision is \emph{independent} of the positions of the particles, so $\langle \vec{r}_{\alpha\beta} \rangle = 0$ and $\langle \Delta \vec{v}_\alpha \cdot \vec{r}_{\alpha\beta} \rangle = \langle \Delta \vec{v}_\alpha \rangle \cdot \langle \vec{r}_{\alpha\beta} \rangle = 0$ so $\Theta = 0$. 
Since the virial contribution is zero, the equation of state for DSMC is the ideal gas law.
Note that Wagner's proof is for infinitesimal collision cells yet we have the ideal gas law even for finite cell size due to symmetry in the selection of collision partners in DSMC.

Alder found this unsatisfactory, so, together with Frank Alexander, we formulated the Consistent Boltzmann Algorithm (CBA).\cite{alexander1995consistent} 
It is identical to DSMC except that collisions change both the velocities \emph{and} the positions of colliding particles.
Specifically, after computing the post-collision velocities the positions are shifted as
\begin{equation}
    \vec{r}_\alpha \Rightarrow \vec{r}_\alpha + d \,\vec{\hat{a}}_\alpha
    \qquad ; \qquad
    \vec{r}_\beta \Rightarrow \vec{r}_\beta - d \,\vec{\hat{a}}_\alpha
\end{equation}
where $\vec{\hat{a}}_\alpha = \Delta \vec{v}_\alpha / |\Delta \vec{v}_\alpha|$ is a unit vector in the direction of the apse line (line between centers) for a hard sphere collision with this change in velocity.
Simulations verified that CBA gives the correct pressure and reasonably accurate transport coefficients up to moderately high densities.

The Consistent Universal Boltzmann Algorithm (CUBA) makes the post-collision displacement a function of density and temperature, which allows one to choose the desired equation of state.\cite{alexander1997CUBA,hadjiconstantinou2000CUBA}
For example, by choosing the van der Waals equation, we were able to simulate the condensation of vapor into liquid droplets.
However, for simulating liquids, we found that CUBA was not competitive with MD in terms of computational cost.

Today, the most popular variants of DSMC for dense gases are based on the Enskog equation.\cite{frezzotti1997Enskog,montanero1997Enskog}
Instead of using a CBA displacement, the correct virial $\Theta$ can be obtained by using a collision probability that depends on both the relative velocity and the relative position for a collision pair.\cite{donev2008densegas}
In other words, the Enskog perspective is that particle positions affect collisions, while in CBA, collisions affect particle positions.

\section{A lesson on transport from molecular dynamics}

In his first book, Graeme Bird said of DSMC, ``the results are remarkably insensitive...and no deleterious effects are generally present...if the cell size approaches the mean free path or if the time step approaches the mean collision time.''\cite{bird1976molecular}
Nevertheless, the demonstration program in this book for the Rayleigh problem sets the collision cell size to half a mean free path and the time step to a fifth of a mean free collision time.

A common way of quantifying the error due to cell size and time step is to measure viscosity or thermal conductivity and, at low Knudsen number, compare with Chapman-Enskog theory.
These transport coefficients can be obtained by measuring the momentum (or heat) flux in a DSMC simulation with a linear velocity (or temperature) gradient.\cite{bird2009accuracy}
From the perspective of non-equilibrium thermodynamics, the transport coefficient is an Onsager coefficient that relates the thermodynamic flux to the thermodynamic force due to the gradient.\cite{deGroot2013non,ottinger2005beyond} 

Yet in the early days of molecular dynamics, such non-equilibrium measurements were not feasible since simulations were limited to a few hundred particles.
Instead, transport properties were measured in simulations of equilibrium systems using results from statistical mechanics.

The basic idea is to apply the fluctuation-dissipation theorem, which connects statistical fluctuations to the rate of dissipation. 
The best known example is Brownian motion in which the mobility, $\mu$, of a particle is obtained from measuring the variance of its position in time as
\begin{equation}
    \mu = \frac{1}{6kT} \lim_{t\rightarrow\infty} 
    \frac{\langle  \left| \vec{r}(t) - \vec{r}(0)\right|^2 \rangle}{t}
\end{equation}
By contrast, a non-equilibrium measurement obtains the mobility $\mu = v_\mathrm{T}/F_\mathrm{ext}$ from the terminal velocity, $v_\mathrm{T}$, for the particle under an external force of magnitude $F_\mathrm{ext}$.

Using Einstein-Helfand theory one can measure the viscosity of a fluid from molecular trajectories as~\cite{alder1970studies}
\begin{equation}
     \eta = \frac{1}{2VkT} \lim_{t\rightarrow\infty} 
    \frac{\langle  \left[ G(t) - G(0)\right]^2 \rangle}{t}
    \qquad \mathrm{with} \qquad
    G(t) = m \sum_{i=1}^N \dot{x}_i(t) y_i(t)
\end{equation}
where $N$ is the number of molecules, $V$ is the system volume, $x_i$, $y_i$ are components of $\vec{r}_i$, and $\dot{x}\equiv dx/dt$.
This allows us to measure the viscosity (and other transport coefficients) from \emph{equilibrium} simulations. A related approach uses the off-diagonal stress tensor component,
\begin{equation}
    J(t) = \dot{G}(t) = m \sum_{i=1}^N \dot{x}_i(t) \dot{y}_i(t)
    + \frac12 \sum_{i=1}^N \sum_{i=j}^N F_{ij}^x (y_i - y_j)
    \label{eqn:GreenKuboJ}
\end{equation}
where $\vec{F}_{ij}$ is the force between molecules. 
Our previous result now takes the more familiar form from Green-Kubo theory,
\begin{equation}
    \eta = \frac{1}{VkT} \int_0^\infty \langle J(t) J(0) \rangle \,dt
    \label{eqn:GreenKuboEta}
\end{equation}
Notice from (\ref{eqn:GreenKuboJ}) that $J$ is the sum of a kinetic (ballistic) contribution and a contribution due to the transfer of momentum by intermolecular forces.

For hard sphere collisions Wainwright wrote~\cite{wainwright1964calculation}
\begin{equation}
    J(t) = m \sum_{i=1}^N \dot{x}_i(t) \dot{y}_i(t) 
    + \sum_c^\mathrm{collisions} \frac{m}{d^2} (\Delta \vec{v}_\alpha \cdot \vec{r}_{\alpha\beta})
    (x_\alpha - x_\beta)(y_\alpha - y_\beta) \delta(t - t_c)
\end{equation}
where $t_c$ is the time of a collision.
He was able to evaluate (\ref{eqn:GreenKuboEta}) and express it as the sum of three contributions, $\eta = \eta^\mathrm{K} + \eta^{\mathrm{K}\times\mathrm{C}} + \eta^\mathrm{C}.$
The kinetic contribution $\eta^\mathrm{K}$ is precisely the Chapman-Enskog viscosity.
The cross term $\eta^{\mathrm{K}\times\mathrm{C}}$ and the collision term $\eta^\mathrm{C}$ are dense gas corrections.

In DSMC, the collision term $\eta^\mathrm{C}$ is \emph{not} zero due to the finite distance between the collision partners (i.e., the finite collision cell size).
Unlike the equation of state contribution due to the virial (see (\ref{eqn:virial})), this contribution to the viscosity does not average to zero over DSMC collisions within a cell.
In fact, following a similar analysis as Wainwright one finds
\begin{equation}
    \eta^\mathrm{C} = \frac{m^2 \Gamma}{2 k T} \left\langle (y_\alpha - y_\beta)^2 \right\rangle
    \left\langle (\Delta\dot{x}_\alpha - \Delta \dot{x}_\beta)^2 \right\rangle
\end{equation}
For cubic collision cells of size $\ell$ the average $\left\langle (y_\alpha - y_\beta)^2 \right\rangle = \ell^2/6$; at equilibrium $\left\langle (\Delta\dot{x}_\alpha - \Delta \dot{x}_\beta)^2 \right\rangle = 4kT/3m$.
In DSMC the cross term $\eta^{\mathrm{K}\times\mathrm{C}}$ is zero and so the hard sphere viscosity is~\cite{alexander1998cell}
\begin{equation}
    \eta = \eta^\mathrm{K} + \eta^\mathrm{C} 
    = \frac{5}{16 d^2} \sqrt{\frac{mkT}{\pi}} \left( 1 + \frac{16}{45\pi} \frac{\ell^2}{\lambda^2}\right)
\end{equation}
where $\lambda$ is the mean free path.
Note that the error goes as $\ell^2$; for a cell size of one mean free path the viscosity deviates from its Chapmann-Enskog value by 11 percent.
A similar analysis (with a similar result) may be applied to obtain the error in thermal conductivity due to finite cell size.
Finally, this approach also gives the error in the transport coefficients due to a finite time step $\tau$;
in DSMC this error goes as $\tau^2$.\cite{garcia2000time,hadjiconstantinou2000analysis}

In his last book Bird wrote ``(In) early DSMC programs... the mean spacing between collision pairs was too large in comparison with the mean free path and this led to the introduction of sub-cells...(and) an option to select the collision pairs from the nearest-neighbour pairs within the cell.''\cite{bird2013dsmc}
He also mentions that ``The time step is set to a specified small fraction of the sampled mean collision time.''
A challenge for modern DSMC codes is to minimize the distance between colliding particles and to minimize the time between ballistic movement steps while avoiding unphysical bias errors (see Section~\ref{sec:demons}).

\section{Fluctuations -- One man's garbage...}

My introduction to DSMC came in 1983 as I was finishing my doctoral studies at The University of Texas at Austin in what is now The Ilya Prigogine Center for Studies in Statistical Mechanics and Complex Systems. At that time there were theoretical results, confirmed by light scattering experiments, indicating that hydrodynamic fluctuations had weak, long-ranged correlations in non-equilibrium systems (e.g., under a temperature gradient).

In his first book, Bird describes fluctuations in DSMC as an unavoidable annoyance but that with a sufficiently large sample size the desired mean value can be measured to the desired accuracy.\cite{bird1976molecular}
One commonly held misconception was that these fluctuations came from the Monte Carlo nature of the collision algorithm in DSMC. 
Yet the very same fluctuations are also found in molecular dynamics, which has deterministic dynamics.

Due to computer memory constraints, the number of particles (or ``simulators'') in a DSMC calculation is typically a small fraction of the number of physical molecules.
For example, at standard temperature and pressure, the mean free path in air is roughly 60~nanometers and a cubic mean free path contains about 6000 molecules. 
In contrast, a typical DSMC simulation uses 20 particles per cubic mean free path, that is, each DSMC particle represents $F_N = 300$ physical molecules.
This number scales as the Knudsen number so it is much higher for rarefied flows.

Statistical mechanics gives us the standard deviation of hydrodynamic variables, allowing us to estimate the expected statistical standard error in particle simulations.\cite{hadjiconstantinou2003statistical}
For example, in a cell with $N_\mathrm{c}$ particles, the variance in each component of fluid velocity is $\langle \delta u_x^2 \rangle = kT/m N_\mathrm{c}$.\cite{landau2013statistical} 
For both MD and DSMC the fractional error in estimating $u_x$ from $S$ independent statistical samples is
\begin{equation}
    E = \frac{\sqrt{\langle \delta u_x^2 \rangle}/\sqrt{S}}{|u_x|}
    \approx \frac{1}{\sqrt{S N_\mathrm{c}}}\,\frac{1}{\mathrm{Ma}}
\end{equation}
where $\mathrm{Ma}$ is the Mach number of the flow.
This result illustrates why nanofluidic flows, in which $\mathrm{Ma} \ll 1$, are much harder to measure in DSMC compared to hypersonic flows.
In DSMC $N_\mathrm{c}$ is smaller than the number of physical molecules by a factor of $F_N$, the variance in fluid velocity increases by the same factor. 

After accounting for this amplification factor, we find that hydrodynamic fluctuations in DSMC are physically correct \cite{garcia1991pop} and yield hydrodynamic information.
For example, from the time correlation of density fluctuations one may measure the sound speed and viscosity from the location and width of the Brillouin peak, mimicking the technique used by light scattering.\cite{baras1995particle,bruno2017rayleigh}
This dynamic structure factor is a useful way to compare DSMC models (e.g. internal energy relaxation) with laboratory measurements and theoretical predictions.\cite{bruno2022internal}

The first numerical observations of long-ranged correlations of hydrodynamic fluctuations were made using DSMC for systems with a temperature gradient \cite{mansour1987fluctuating} and a velocity gradient \cite{garcia1987hydrodynamic}.
The ``giant fluctuation'' phenomenon, first reported in diffusive mixing experiments \cite{vailati1997giant}, is due to such non-equilibrium correlations for fluctuations of concentration and fluid velocity, as confirmed by DSMC \cite{donev2011diffusive}.
Most recently DSMC simulations have been used to study the impact of thermal fluctuations on turbulence.\cite{gallis2017molecular,mcmullen2022navier} 
They confirm theoretical predictions~\cite{bandak2022dissipation} and numerical fluctuating hydrodynamic observations~\cite{bell2022thermal} that  
molecular fluctuations \emph{dominate} the turbulent energy cascade at the Kolmogorov length scale.

Bird wrote in his last book,
``While the fluctuations are unphysical when $F_N$ is large, they are physically realistic... (with) a one-to-one correspondence between real and simulated molecules. This is another instance of DSMC going beyond the Boltzmann equation because fluctuations are neglected in the Boltzmann model.''
His last three papers were on Brownian motion.\cite{bird2015rate,bird2016brownian,bird2017possibility}

\section{Exorcising the demons}\label{sec:demons}

In 1887 Maxwell presented a thought experiment,
``... conceive of a being whose faculties are so sharpened that he can follow every molecule in its course ...
so as to allow only the swifter molecules to pass from (chamber) A to B, and only the slower molecules to pass from B to A. 
He will thus, without expenditure of work, raise the temperature of B and lower that of A, in contradiction to the second law of thermodynamics.''
The DSMC algorithm has full information and control of the particle's dynamics and modern implementations have many complex stages.
How can we be sure that we do not have a Maxwell demon hiding in our simulations?
In other words, how do we test that our simulations satisfy the Second Law of Thermodynamics?
Boltzmann's H-theorem is not helpful, since, due to fluctuations, it only applies in the limit $N \rightarrow \infty$.

Consider an isolated system with microstate dynamics\cite{thomsen1953logical}
\begin{equation}
    \dot{p}_i = \sum_j^\mathrm{states} ( p_j W_{ji} - p_i W_{ij} )
\end{equation}
where $p_i$ is the probability of state $i$ and $W_{ij}$ is the transition rate from $i$ to $j$.
At equilibrium $\dot{p}_i = 0$.
If $W_{ij} = W_{ji}$, then we have \emph{microscopic reversibility}.
At equilibrium we have \emph{detailed balance} ($p_j W_{ji} = p_i W_{ij}$) and \emph{ergodicity} ($p_i = p_j$) iff we have microscopic reversibility.
Finally, the dynamics satisfies the Second Law ($\dot{S} = - k \sum_j \dot{p}_j \ln p_j \geq 0$) iff we have ergodicity.\cite{thomsen1953logical}

Detailed balance is a necessary condition for microscopic reversibility, which is a sufficient condition for the dynamics to obey the Second Law.
Yet, it is not uncommon to find DSMC implementations that do \emph{not} satisfy detailed balance.
For example, some of the early surface scattering models were ad hoc empirical fits to experimental data, which led Cercignani and Lampis to formulate a model that imposed detailed balance.\cite{cercignani1971kinetic}
Many implementations of inflow/outflow boundaries also fail to satisfy detailed balance.\cite{tysanner2005non}
If machine learning methods are used to handle some elements in a DSMC calculation, such as chemical reactions and internal degrees of freedom, we may not know if detailed balance is satisfied.\cite{ball2024online}
And even when the algorithm in a DSMC code is theoretically sound, there is always the possibility that the implementation has a 'bug' (coding error).

In testing for the presence of Maxwell demons in a DSMC code, equilibrium hydrodynamic fluctuations can serve as a coal mine canary.
For example, at equilibrium, the number of particles in each cell is independently Poisson distributed, and
this result is independent of the collision model, boundary conditions, etc.
Statistical mechanics provides further relations for the variances and correlations of fluid velocity, temperature, internal energy, and other variables.\cite{garcia2007estimating,landau2013statistical}
While this is a necessarily but not sufficient statistical test, it is a sensitive one that can reveal malevolent problems.
Ensuring the correctness of simulation data is increasingly important as we enter the age of artificial intelligence. 
Given the AI systems' voracious appetite for such data, we must avoid having machines learn the wrong lessons. 

\section{Concluding Remarks}
Graeme Bird and Berni Alder met for the first time at the 22nd Rarefied Gas Dynamics Symposium in Sydney, Australia, nearly 40 years after their original publications on DSMC and MD.
In the 25 years since that meeting, we find increasing cooperation between the two algorithms' communities. 
One example is the study of rotational relaxation using all-atom molecular dynamics simulations combined with DSMC.\cite{valentini2012molecular}
Yet perhaps the best example is the shared heritage of LAMPPS~\cite{LAMMPS} and SPARTA~\cite{SPARTA}, two of the most widely used codes for molecular dynamics and direct simulation Monte Carlo.
Although separated at birth, MD and DSMC are now the patriarchs overseeing a large family of particle-based fluid algorithms.

\section*{Acknowledgement}
The author acknowledges support from the US Department of Energy, Office of Science, Office of Advanced Scientific Computing Research, Applied Mathematics Program under contract no. DE-AC02-05CH11231.

%

\bibliography{RGD}  

%
%








\end{document}